\documentclass[conference]{IEEEtran}
\IEEEoverridecommandlockouts
\usepackage{multirow}
\usepackage{float}
\usepackage{cite}
\usepackage{amsmath,amssymb,amsfonts}
\usepackage{algorithmic}
\usepackage{graphicx}
\usepackage{textcomp}
\usepackage{xcolor}
\usepackage{comment}
\renewcommand\IEEEkeywordsname{Keywords}
\def\BibTeX{{\rm B\kern-.05em{\sc i\kern-.025em b}\kern-.08em
    T\kern-.1667em\lower.7ex\hbox{E}\kern-.125emX}}
\begin{document}

\title{ A detailed study on novel metamaterial absorber at WR3 band\\

}

\author{\IEEEauthorblockN{Sarvesh Gharat}
\IEEEauthorblockA{\textit{Department of Electronics and Telecommunication} \\
\textit{Vishwakarma Institute of Information Technology}\\
Pune, India \\
sarveshgharat19@gmail.com}
\and
\IEEEauthorblockN{Shriganesh Prabhu}
\IEEEauthorblockA{\textit{Department of Condensed Matter Physics} \\
\textit{Tata Institute of Fundamental Research}\\
Mumbai, India \\
prabhu@tifr.res.in}
}

\maketitle

\begin{abstract}
Frequency band of 30 to 300 GHz, which is termed as mm wave band offers 100 times bandwidth as compared to sub 6 GHz band. In this study, we propose a novel design which acts as a near unity absorber at 258 GHz or 0.258 THz in Y polarisation and 292 GHz or 0.292 THz in X polarisation. The study also includes effect of change in parameters on resonance frequency in both X and Y polarisation which allows us to use same design with modified parameters to get near unity absorptivity in frequency of 205 to 345 GHz. Further, we also discuss on higher frequency mode which particularly appears in Y polarisation and on changing the inner square dimension of the design.
\end{abstract}





\section{Introduction}

There are some properties such as negative permitivity and permeability which are absent in materials which are present in nature. Introducing metamaterials helps us to achieve all these properties \cite{kshetrimayum2004brief}\cite{capolino2017theory}\cite{walser2000metamaterials}. This unique properties of metamaterials are mainly observed at it's resonant frequencies \cite{bukhari2019metasurfaces}.\\

Metamaterials can be treated as an LC circuit. Both inductance and capacitance depends on structure of metamaterial, polarisation of incident wave, thickness of substrate and refractive index of substrate used \cite{sangala2020single}\cite{chowdhury2011broadband}\cite{tao2008highly}. Although, we are familiar with variation of these properties for specific structures \cite{marwaha2016accurate}\cite{george2018mathematical}\cite{elander2011mathematical}\cite{bose2012mathematical}, we don't have any universal theoretical model helping us to get the exact relations. \\

Use of metamaterials is widely done at many places such as sensor identification, high-frequency battlefield communication, frequency filter and many more \cite{valipour2021metamaterials}\cite{rappaport2019wireless}\cite{singh2015review}. One such application is using it as a perfect unity absorber. These metamarial absorbers are further widely used is applications such as thermal emitters, photovoltaic cells, optical imaging devices, etc \cite{diem2009wide}
\cite{munday2011large} \cite{hu2010plasmonic}. The first metamaterial absorber was experimentally confirmed by Landy et al. \cite{landy2008perfect} in 2008. Since than, a lot of work has been done in creating novel metamaterial absorbers  \cite{wang2020multiple}\cite{gandhi2021ultra}\cite{liu2020ultrathin}\cite{luo2021multiband}\cite{elakkiya2020seven} to be used in different applications as stated before. \\

In today's era where telecommunication industry is developing exponentially, it is important to also have near unity absorbers in different bands of 6G \cite{chi2020progresses}\cite{li2021recent}. In our previous work \cite{https://doi.org/10.48550/arxiv.2203.13442}, we had propose a novel metamaterial absorber in WR 1 and WR 1.5 bands. However, there is a strong need to have near unity absorbers in remaining bands of 6G. Hence, in this study we propose a novel metamaterial structure which will act as a near unity absorber in specific frequencies of WR 3 band. Along with that, we also provide a study on change in central frequencies due to variation in Geometrical parameters and change in polarisation. 

\section{Design and Simulations}
This section focuses on schematic of proposed design. Along with that, this section will also includes the necessary boundary conditions and assumptions considered in this study.
\begin{figure}[hbt!]
  \includegraphics[width = 8.5 cm]{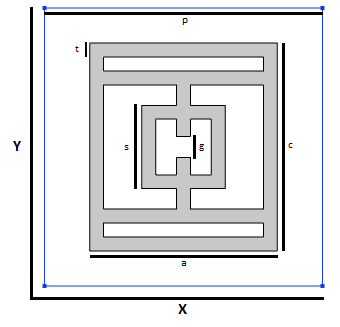}
  \caption{Proposed Design}
  \label{fig:fig1}
\end{figure}
The proposed design as shown in Figure \ref{fig:fig1} has symmetry along both X and Y axis. The proposed design is simulated over Kapton with thickness of 25$\mu$m having refractive index of $1.88 + 0.04i$. 

\begin{table}[ht]
\begin{center}
\begin{tabular}{|c|c|c| } 
\hline
 Parameter & Value\\
\hline
p & 300 \\ 
\hline
a & 202.5 \\ 
\hline
c & 224 \\ 
\hline
s & 90 \\
\hline
g & 22.5 \\ 
\hline
t & 15 \\ 
\hline
\end{tabular}
\end{center}
\caption{Value of Parameters in $\mu$m as shown in Figure\ref{fig:fig1}}
\label{table:1}
\end{table}

All the values of parameters as shown in Figure\ref{fig:fig1} can be seen in Table\ref{table:1}. The parameters given in Table\ref{table:1} happen to be optimized parameters to get resonant frequency of 235~GHz.\\

To simulate the proposed design, use of COMSOL MULTIPHYSICS \textregistered, is done. Utmost care is taken while considering boundary conditions so as to reduce the required computational power. The used conditions are elaborated below:
\begin{itemize}
    \item Perfectly Electric Conductor: It is a special case of electric field boundary condition that sets tangential component of electric field to zero. It is generally used in modelling a lossless metallic surface \cite{multiphysics1998introduction}.
    \item Periodic Condition: Periodic condition is used to make unit cell structure periodic along particular direction \cite{multiphysics1998introduction}. 
\end{itemize}
In the proposed simulation, the structure is periodic along both X and Y direction. Along with that, use of Perfectly Electric Condition (PEC) condition is done on the design as well as the end of the substrate. Adding PEC boundary condition at the bottom of the substrate allows us to omit air domain after substrate helping us to reduce the computational power.\\

To vary the parameters as listed in Table\ref{table:1}, use of parametric sweep is done. The parametric sweep feature as provided by COMSOL\cite{multiphysics1998introduction} helps us to iterate the process of finding a solution over complete set of variable. This, not only helps us in reducing the manual work but also helps in comparing the results at one place. Table\ref{table:2} tells us about the range in which individual parameters are varied.

\begin{table}[ht]
\begin{center}
\begin{tabular}{|c|c|c| } 
\hline
 Parameter & Minimum Value & Maximum Value\\
\hline
a & 160 & 260 \\ 
\hline
c & 180 & 280 \\ 
\hline
s & 40 & 140 \\
\hline
g & 5 & 50 \\ 
\hline
\end{tabular}
\end{center}
\caption{Minimum and Maximum values of parameter}
\label{table:2}
\end{table}

Further to calculate absorbance in percentage, we make use of simple formulation between absorbance, transmission and reflection which is
$$\text{i.e Absorptivity } = 1 - \text{abs}(S_{11})^2 - \text{abs}(S_{21})^2$$ where $S_{11}$ and $S_{21}$ are reflection and transmission coefficients. However, as in simulation we make use of PEC at the bottom, the incident wave cant pass through it making $S_{21}$ as 0 and simplifying the formula to $$\text{i.e Absorptivity in }\% = 1 - \text{abs}(S_{11})^2$$ \\

The proposed design is studied in both X and Y polarisation. Further, all the data is extracted in a text file which is further plotted using Origin Pro 2021.

\section{Results and Discussion}
This section comprises of simulated optimised results. Apart from that we will be also discussing on effects of geometrical parameters on Absorbance in both X and Y polarisation.

\subsection{Y Polarisation}

\subsubsection{Optimized Parameters}

\begin{figure}[hbt!]
  \includegraphics[width = 8.5 cm]{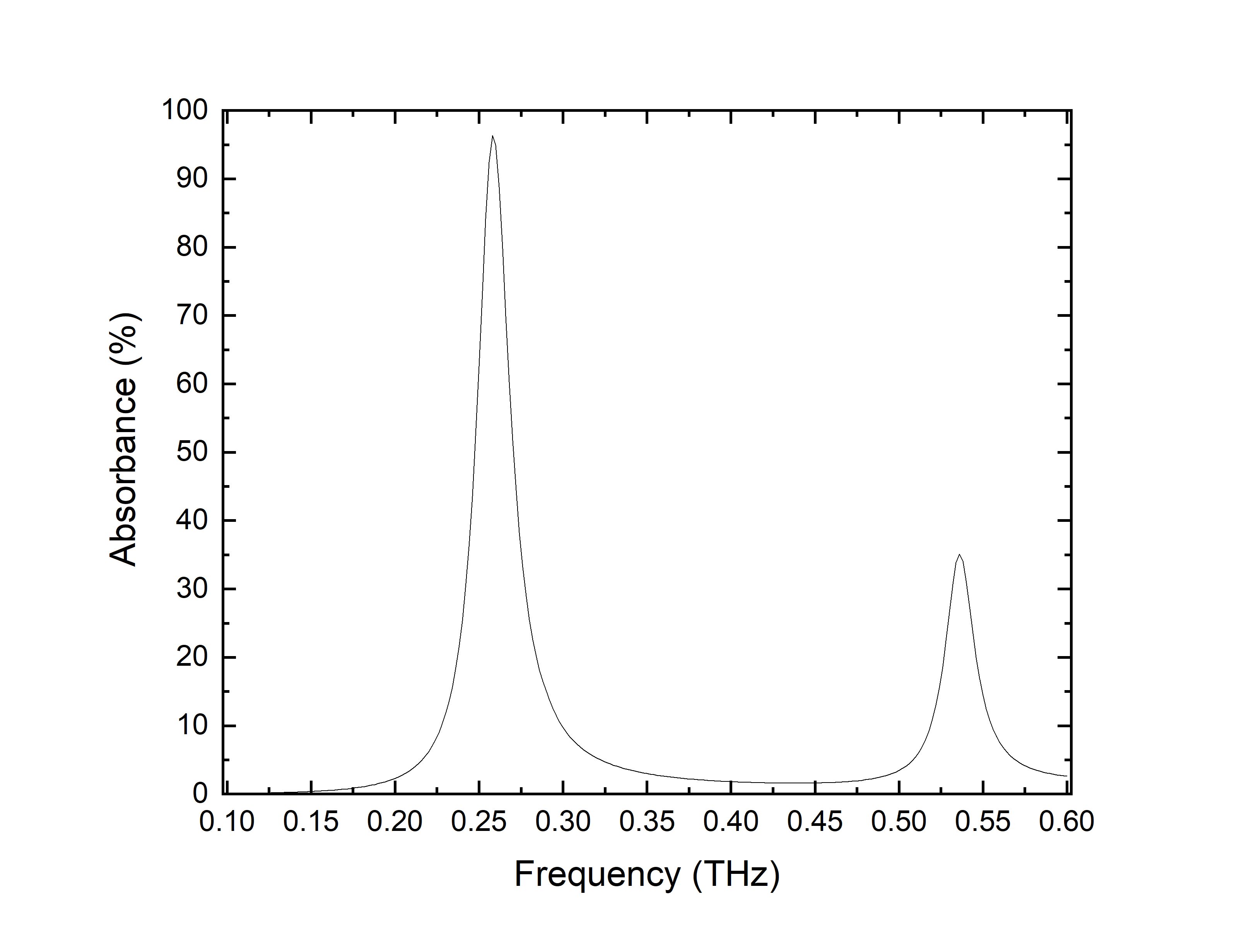}
  \caption{Frequency vs Absorbance plot for Y polarisation}
  \label{fig:fig2}
\end{figure}

As seen in Figure \ref{fig:fig2}, almost 96.3\% of incident EM field is absorbed at frequency of 258 GHz or 0.258 THz. Apart from that there's one more small resonance happening at 0.536 THz with absorptivity of 35.12\%.

\begin{figure}[hbt!]
  \includegraphics[width = 9.5 cm]{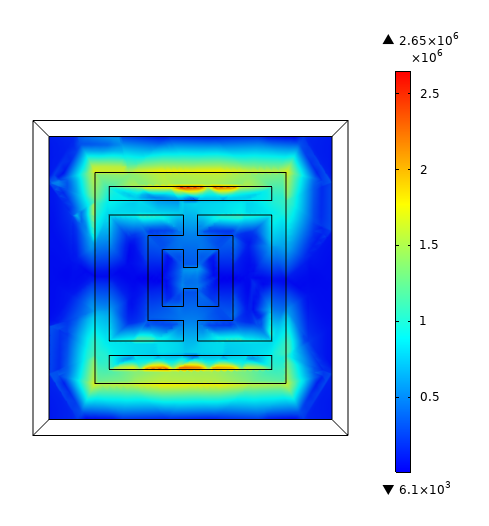}
  \caption{$E_z$ component of Electric field on metamaterial at low frequency mode}
  \label{fig:fig3}
\end{figure}

The initial low frequency mode as seen in Figure \ref{fig:fig2} is mainly due to inductance produced after the confinement of electric field on horizontal arms as shown in Figure \ref{fig:fig3}

\begin{figure}[hbt!]
  \includegraphics[width = 8.5 cm]{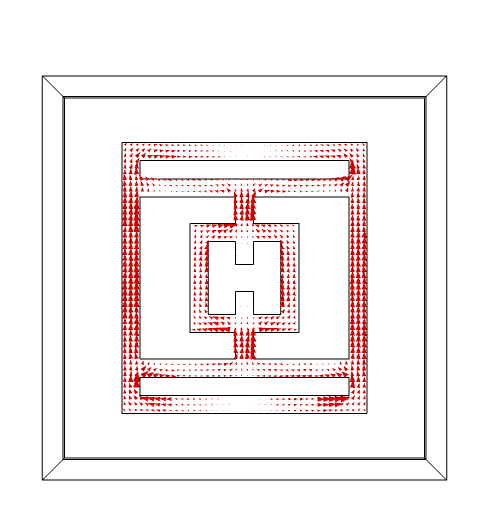}
  \caption{Surface Current Density on metamaterial at low frequency mode in Y polarisation}
  \label{fig:fig4}
\end{figure}

The supporting fact for above happening is evident from Figure \ref{fig:fig4} where we have shown surface current density. 

We see that the current follows a unidirectional path along Y axes. This is consistent with polarisation of Electric field. Apart from that, we also see high current density is both inner and outer vertical arms which leads to attenuation of emitted radiation causing a very high absorbance at low frequency mode.\\

\begin{figure}[hbt!]
  \includegraphics[width = 9.5 cm]{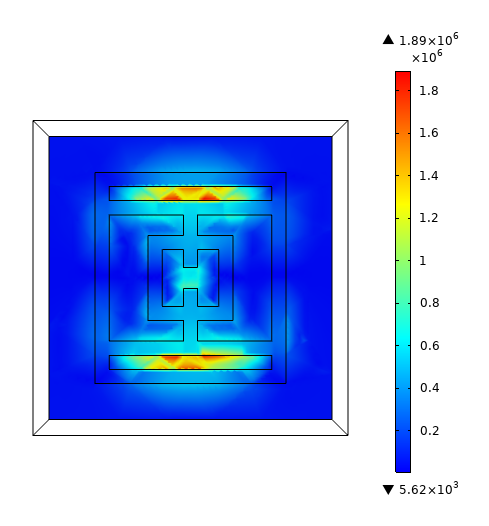}
  \caption{$E_z$ component of Electric field on metamaterial at high frequency mode in Y polarisation}
  \label{fig:fig5}
\end{figure}

On contrary to that the higher frequency mode appears due to creation of capacitance between inner and outer horizontal arms as seen in Figure \ref{fig:fig5}.

\begin{figure}[hbt!]
  \includegraphics[width = 8.5 cm]{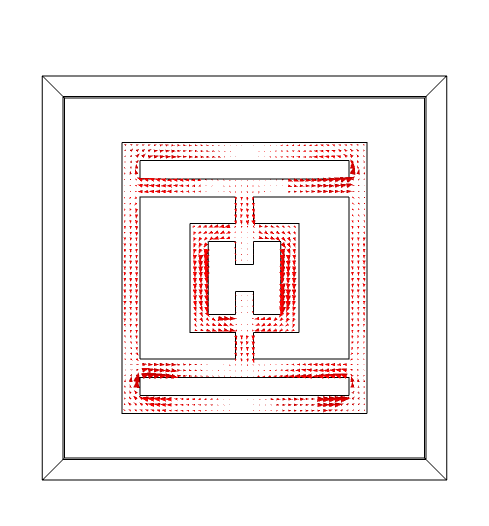}
  \caption{Surface Current Density on metamaterial at high frequency mode in Y polarisation}
  \label{fig:fig6}
\end{figure}

Also in this case, there is flip in direction of current in all vertical arms. This happens to be in opposite direction of the polarisation. With that, the current density is comparatively less than that of what we saw in initial lower frequency mode. This can be concluded by looking into Figure \ref{fig:fig4} and Figure \ref{fig:fig6}

\subsubsection{Variation in Parameters}

To study effect of Geometrical parameters like height (c), width (a), gap (g) and size of square (s) as seen in Figure \ref{fig:fig1}, we vary them with minimum and maximum values as shown in Table \ref{table:2}

\begin{figure}[hbt!]
  \includegraphics[width = 8.5 cm]{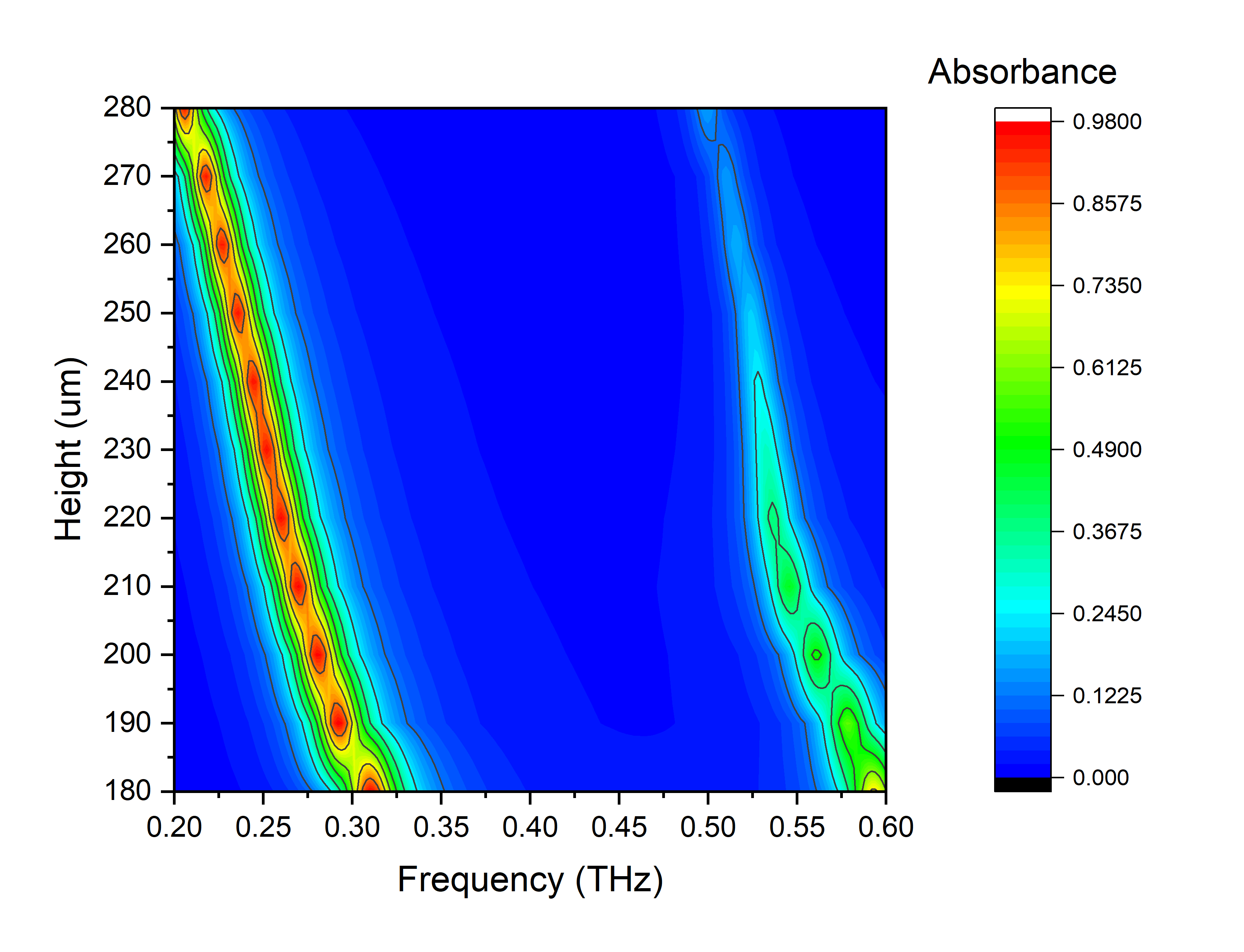}
  \caption{Contour plot of absorbance on varying height in Y polarisation}
  \label{fig:fig7}
\end{figure}

Figure \ref{fig:fig7} refers to variation of resonant frequency on changing height. We see that on increasing the height, the initial mode frequency gets redshifted. As the period remains constant and the height increases, the gap between 2 unit cells decrease resulting into more capacitance and as we know, more is the capacitance, less wil be the resonant frequency. On the other hand, in higher mode, along with redshift we also see decrease in absorbance and at one point  the peak vanishes. This happens because the gap between the upper two parallel arms increases as the length is increased and this reduces the mode that can be sustained within the gap between the two arms. Thus at one point it can become weak enough to vanish.

\begin{figure}[hbt!]
  \includegraphics[width = 8.5 cm]{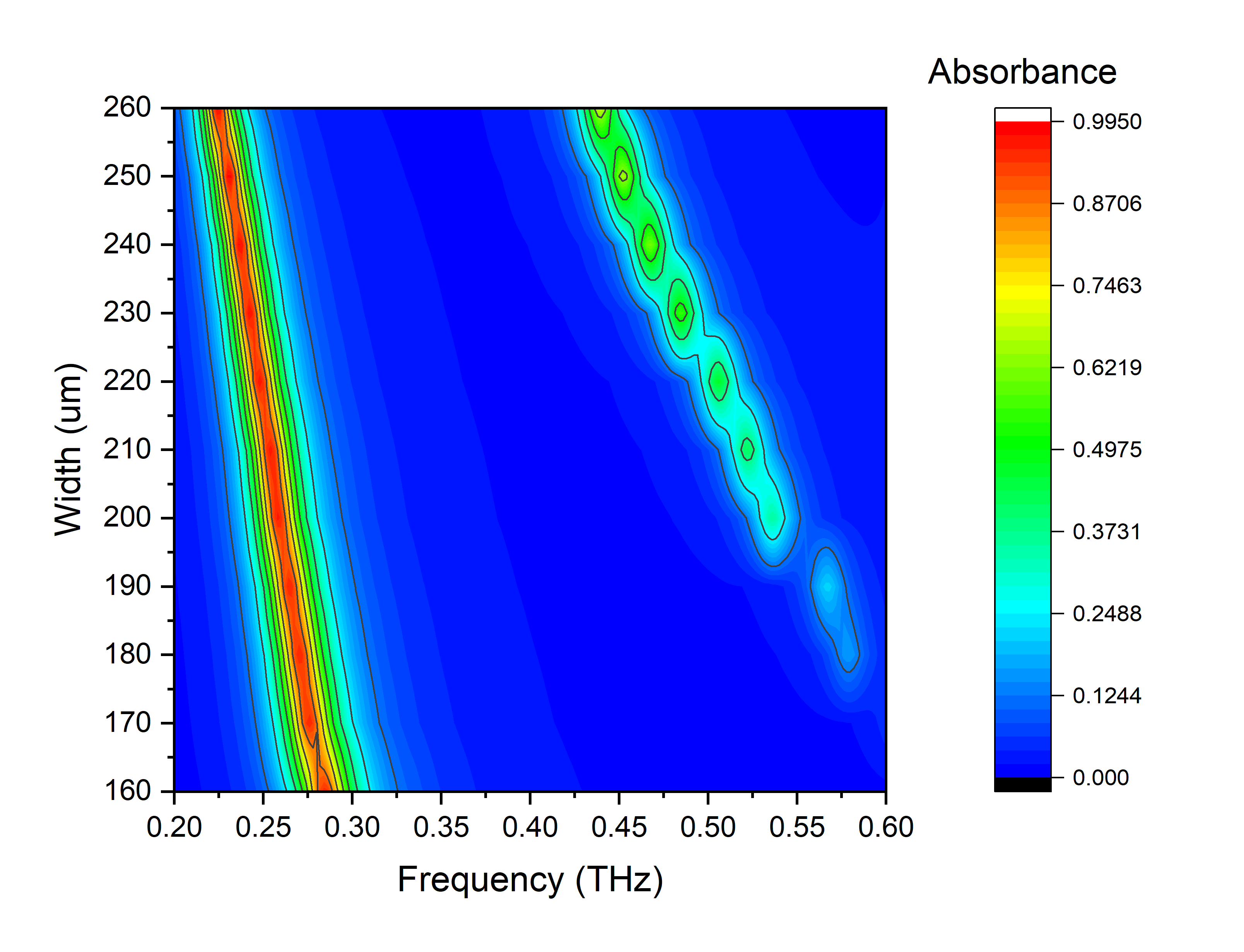}
  \caption{Contour plot of absorbance on varying width in Y polarisation}
  \label{fig:fig8}
\end{figure}

The redshift is also observed on increasing the width as seen in Figure \ref{fig:fig8}. However, in this case the reason for this happening is increase in inductance due to increase in horizontal arm length. On contrary to Figure \ref{fig:fig7}, in Figure \ref{fig:fig8} we see that higher mode vanishes on increasing the width. This can happen as the width is increased, the two side of the cell will start coming closer and then there is a coupling between the two progressively increasing as the width increases. This can result in the appearance of the mode, which is capacitively excited and as the arms draw closer, the capacitance decreases, increasing the frequency. This the mode appears suddenly as the width becomes larger and the mode becomes stronger.

\begin{figure}[hbt!]
  \includegraphics[width = 8.5 cm]{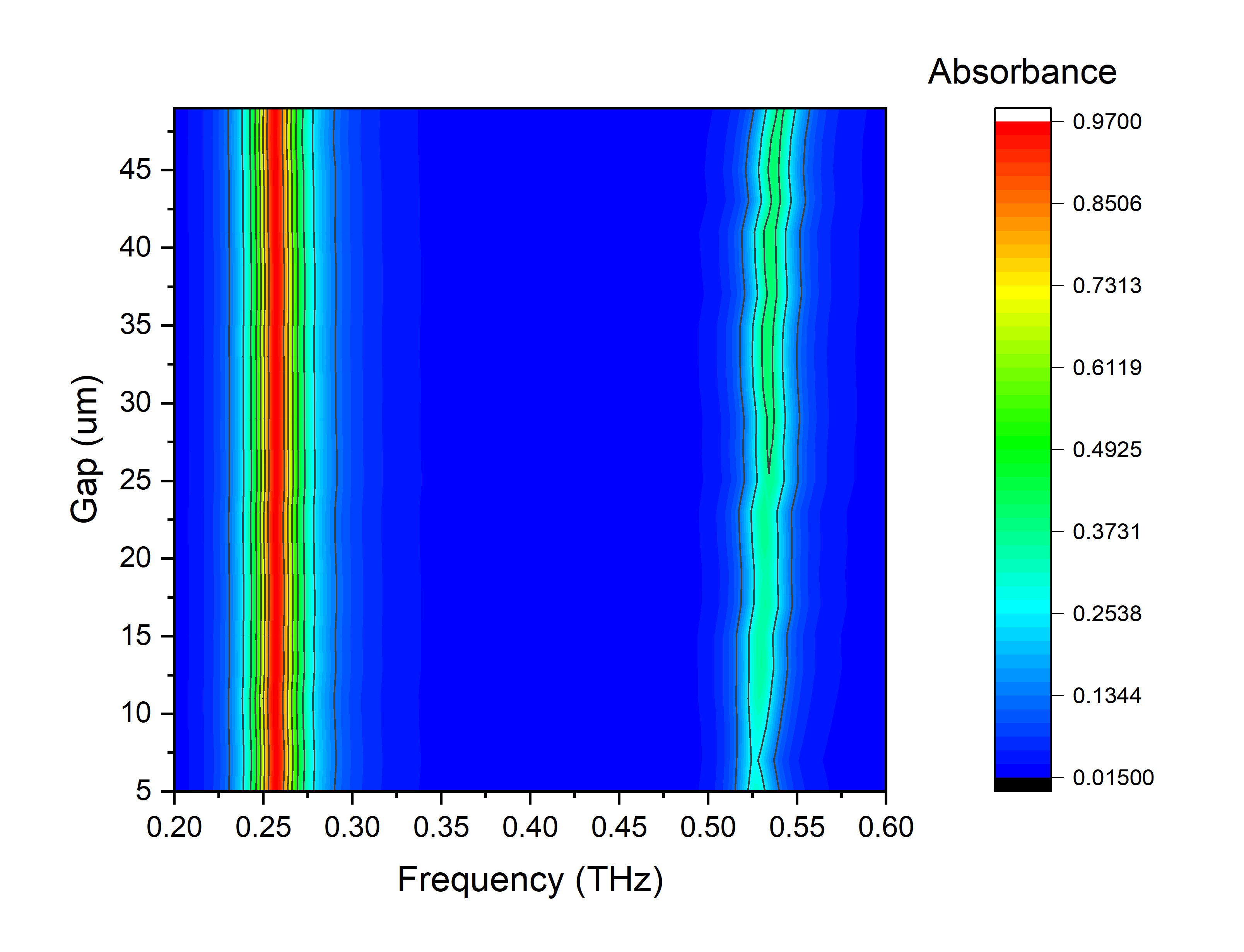}
  \caption{Contour plot of absorbance on varying gap in Y polarisation}
  \label{fig:fig9}
\end{figure}

We further vary gap to understand confirm the production of capacitance in the gap. However, if that was the case then there would had been blueshift in resonant frequency due to decrease in capacitance. But there's no such observation in the initial mode (See Figure \ref{fig:fig9}, hence we can safely say that there's negligible amount of capacitance created in first mode. It is obvious from the figure that when gap will vary, the length remains the same, so frequency should not change. This can also be verified from Figure \ref{fig:fig3} where we don't see any confinement of electric field. On other hand, in high frequency mode, a considerable amount of capacitance is been created as seen in Figure \ref{fig:fig5}. Hence, there's small amount of redshift happening at higher frequency as evident from Figure \ref{fig:fig9}

\begin{figure}[hbt!]
  \includegraphics[width = 8.5 cm]{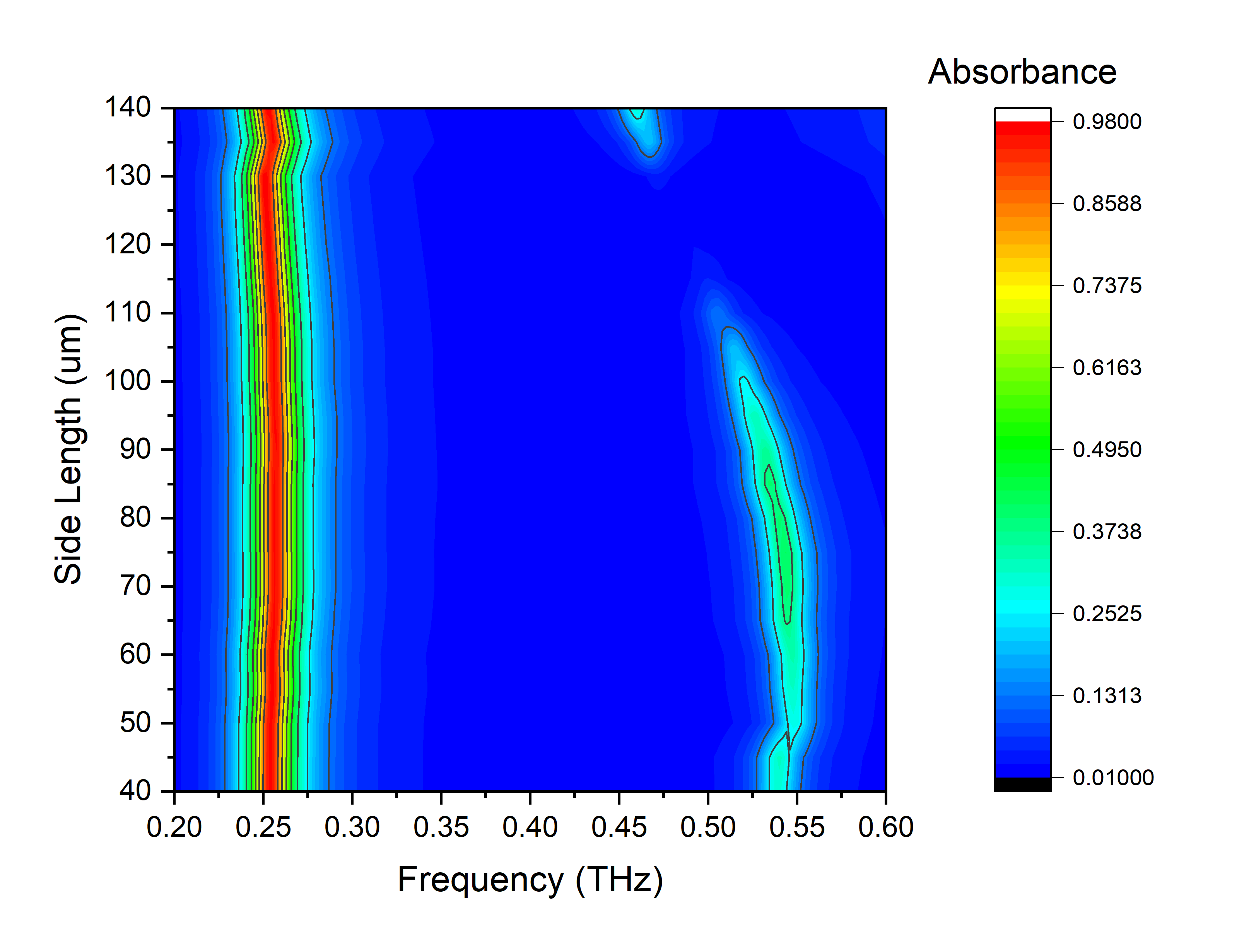}
  \caption{Contour plot of absorbance on varying size of square in Y polarisation}
  \label{fig:fig10}
\end{figure}

From Figure \ref{fig:fig10}, we observe that similar to gap, the internal square plays no major role in initial mode. This is expected as the first mode is due to the upper and lower arm gap capacitance. However, at high frequency, on increasing the inner square size, mode gets redshifted before vanishing away. However it reappears after getting close to exterior loop. This will happen if the inner square , after increasing it's size can start interacting with the upper and lower arm excitation mode. The increasing size of the central portion can result in the modified coupling of the upper-lower arms. So the Frequency start reducing and will vanish at a particular combination of Size and arms length. Now if the size of the inner square increases further, now the capacitance between the upper and lower arms will start dominating, and there will be arise of new frequency modes.Other way to look at it is, the capacitance start increasing with size, decreasing the resonance frequency of the high frequency mode.

\subsection{X Polarisation}

\subsubsection{Optimized Parameters}

\begin{figure}[hbt!]
  \includegraphics[width = 8.5 cm]{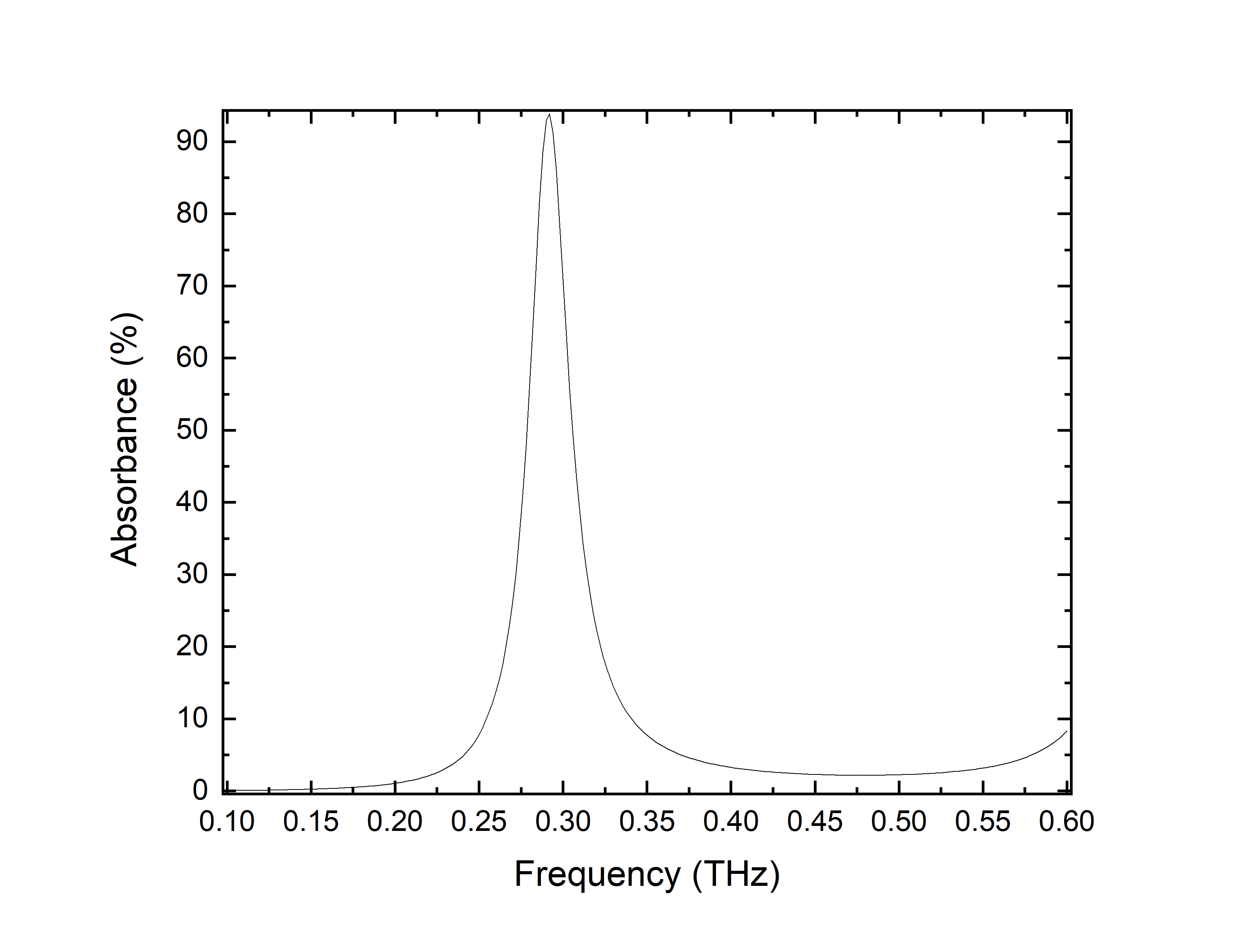}
  \caption{Frequency vs Absorbance plot for X polarisation}
  \label{fig:fig11}
\end{figure}

From Figure \ref{fig:fig11}, we see that on exciting electric field in X direction, there only one mode occurring in bandwidth of 500 GHz at 292 GHz with absorptivity of 93.85 \%.

\begin{figure}[hbt!]
  \includegraphics[width = 9.5 cm]{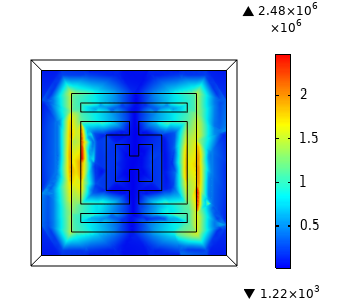}
  \caption{$E_z$ component of Electric field on metamaterial at low frequency mode in X polarisation}
  \label{fig:fig12}
\end{figure}

In this case, the only mode visible is majorly due to the inductance produce in vertical arms as seen in Figure \ref{fig:fig12}. 

\begin{figure}[hbt!]
  \includegraphics[width = 8.5 cm]{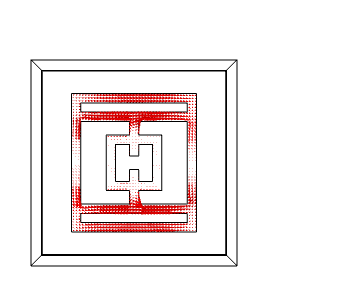}
  \caption{Surface Current Density on metamaterial at low frequency mode in X polarisation}
  \label{fig:fig13}
\end{figure}

The surface current density as observed in Figure \ref{fig:fig13} is unidirectional towards the direction of polarisation. With that, the surface charge density is also negligible in inner square. Apart from that the major reason behind high absorptivity at this frequency is due to attenuation of incident wave because of high current density as seen in Figure \ref{fig:fig13} 

\subsubsection{Variation in Parameters}
We further vary the geometrical parameters as shown in Table \ref{table:2} to study effect of these parameters on absorbance.

\begin{figure}[hbt!]
  \includegraphics[width = 8.5 cm]{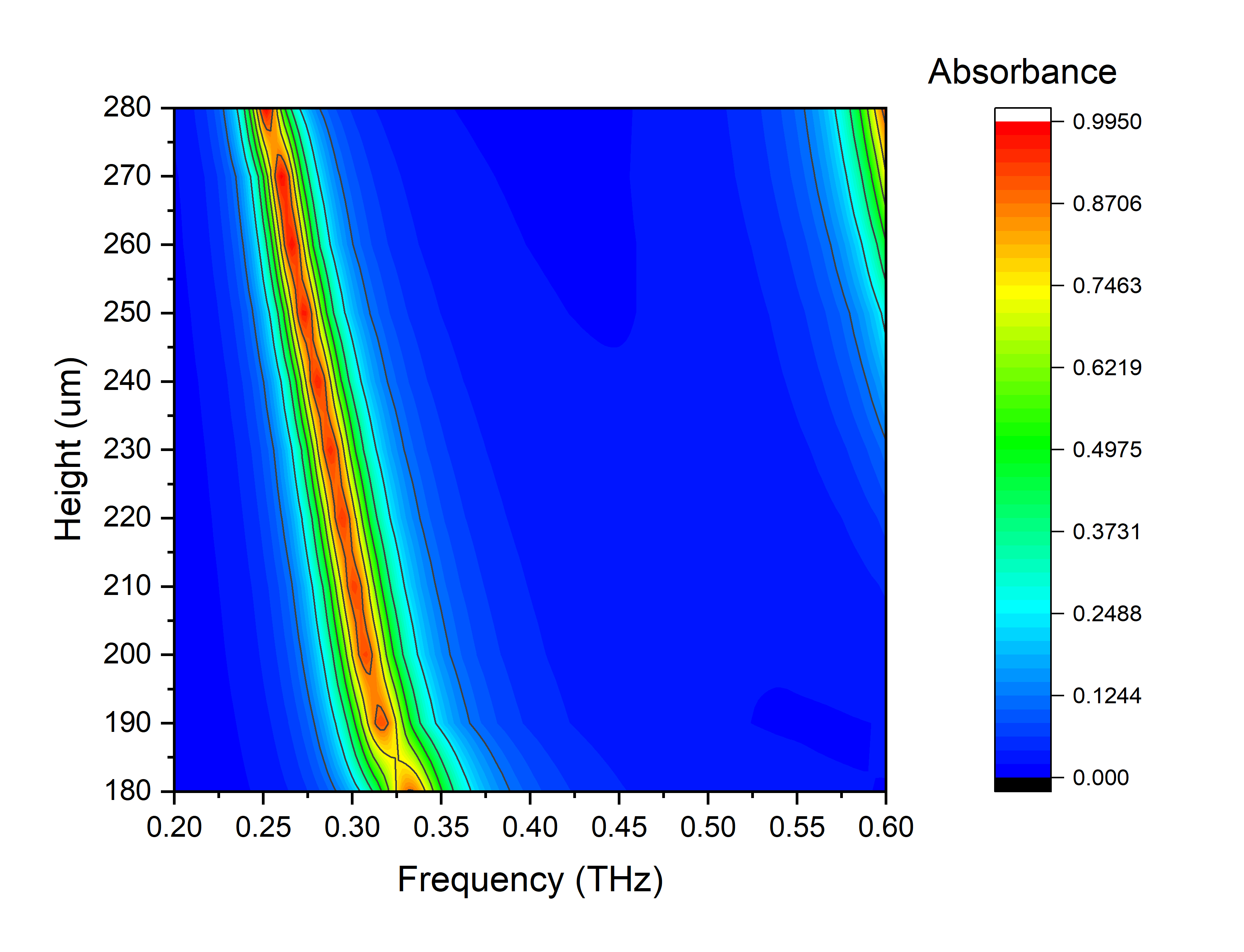}
  \caption{Contour plot of absorbance on varying height in X polarisation}
  \label{fig:fig14}
\end{figure}

Figure \ref{fig:fig14} refers to change in absorptivity on varying the height. For all the values of height considered in this study we have one single mode in frequency from 0.1 to 0.55 THz. However, as we move towards higher frequency there are some instances of higher frequency mode which currently isn't our interest. We also observe a red shift is resonance frequency. This is due to increase in inductance on increasing the arm length. 

\begin{figure}[hbt!]
  \includegraphics[width = 8.5 cm]{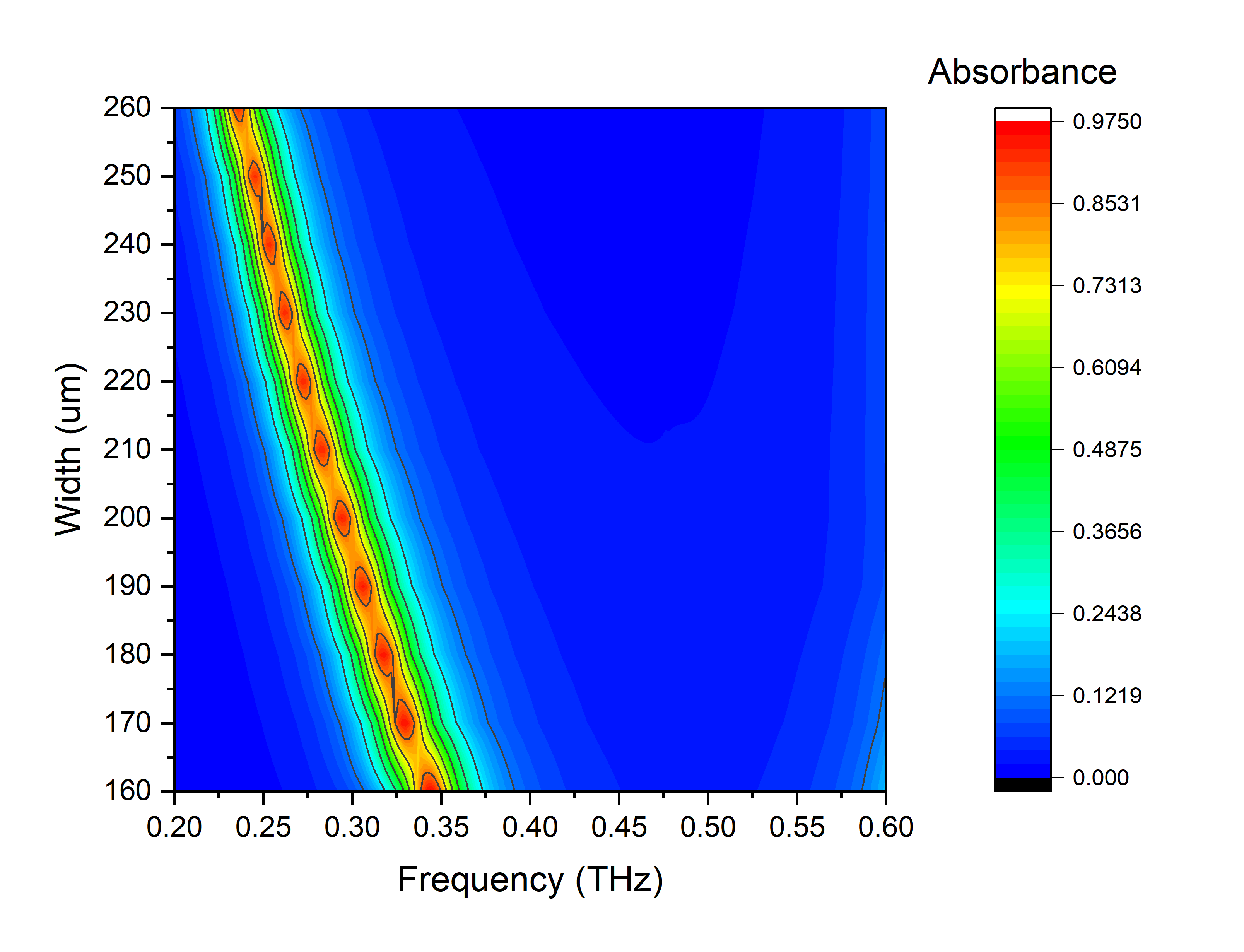}
  \caption{Contour plot of absorbance on varying width in X polarisation}
  \label{fig:fig15}
\end{figure}

\begin{figure}[hbt!]
  \includegraphics[width = 8.5 cm]{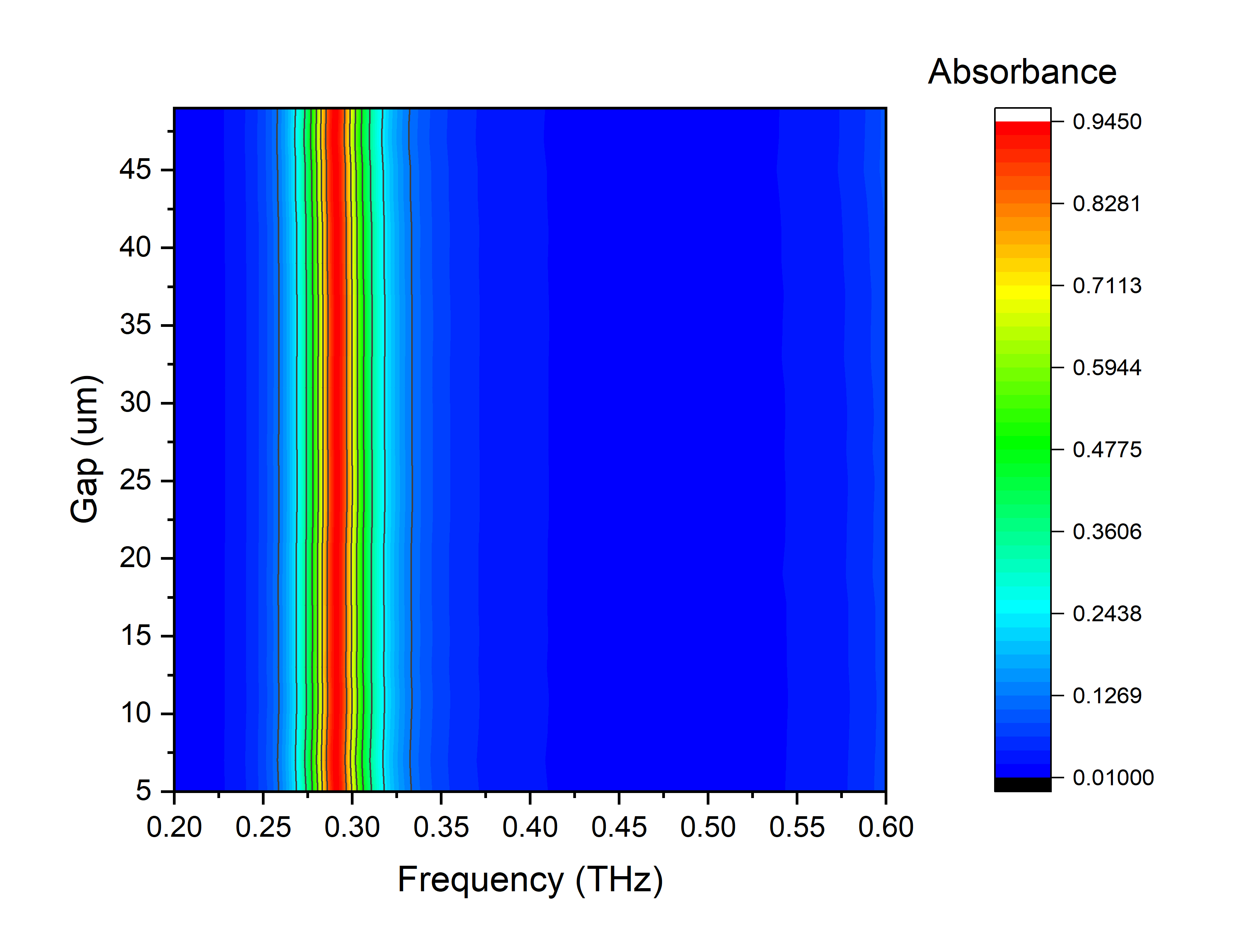}
  \caption{Contour plot of absorbance on varying gap in X polarisation}
  \label{fig:fig16}
\end{figure}

Similarly, on increasing the width we observe a redshift. Here, the redshift occurs mainly because of increase in capacitance which happens due to closeness of exterior arms in every unit cell. Apart from that as in X polarisation, the initial gap doesn't play any role there is no instance of second mode in this case. Th supporting instance for this claim is evident from Figure \ref{fig:fig13} and Figure \ref{fig:fig16}.

In Figure \ref{fig:fig16}, we see that there's no change in resonance frequency on changing the gap size. This happens due to direction of Electric field being parallel to gap. As both the ends of gap will be on equipotential surface in case of X polarisation, theres no capacitance created in this case resulting into no effect of absorbance due to gap.

\begin{figure}[hbt!]
  \includegraphics[width = 8.5 cm]{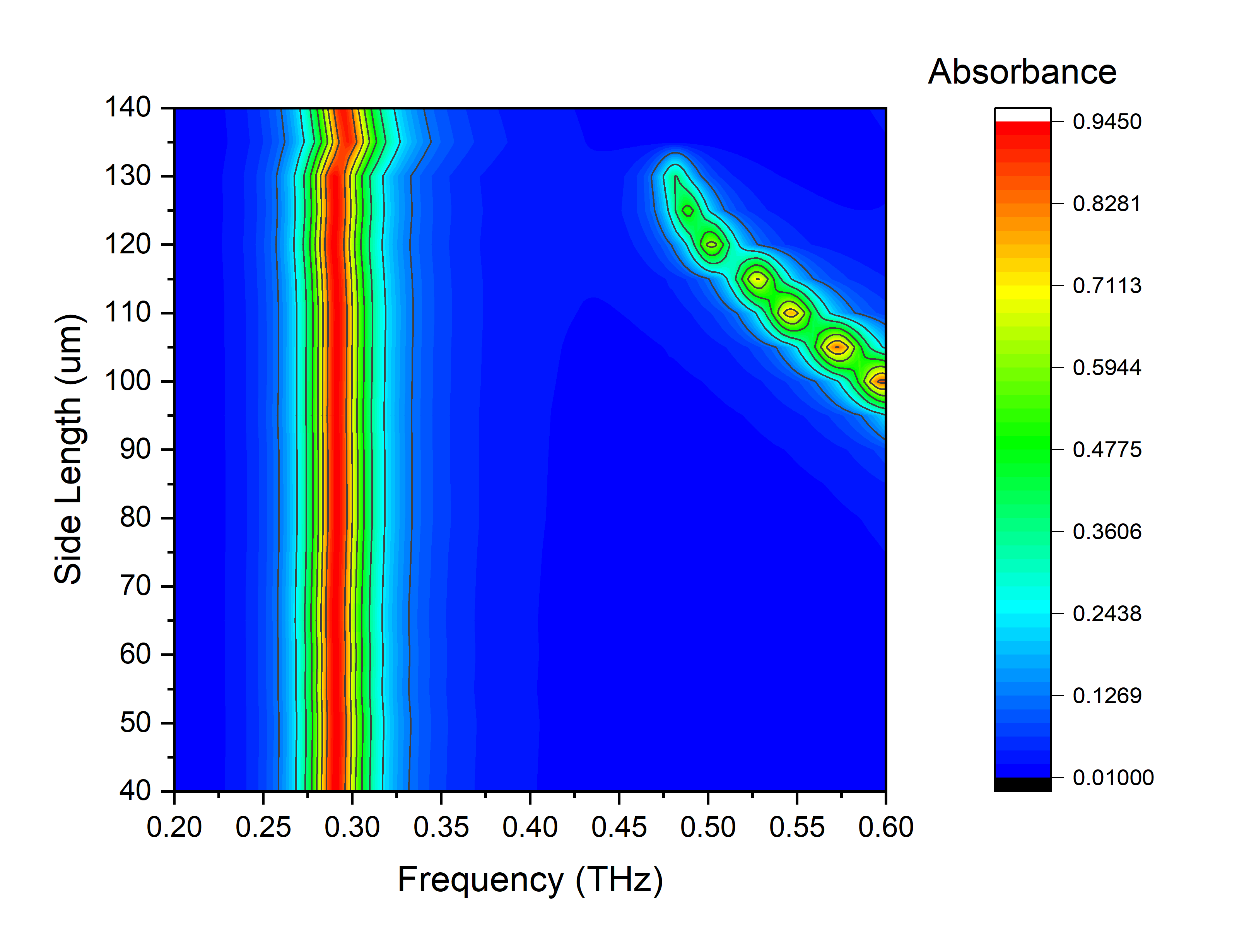}
  \caption{Contour plot of absorbance on varying size of square in X polarisation}
  \label{fig:fig17}
\end{figure}

Similar thing can be observed on changing the size of square as shown in Figure \ref{fig:fig17}, However, in this case we also see a higher mode which starts appearing from 0.5 THz. The mode exists at the higher frequency, but is not visible in our simulations data. However, as the middle square changes it's size, the capacitive coupling must be getting reduced, increasing the capacitance, decreasing the resonance frequency. This trend is clearly seen in the figure, where the second mode keeps red shifting, eventually vanishing when the coupling is no more effective.

\section{Conclusion}
In this study, we propose a novel metamaterial structure to be used in 6G band. This design happens to be one of the optimal candidate for 6G applications mainly due to it's low complexity. We further provide a study on effect on resonance frequency on changing of geometric parameters like width, height, gap and size of square in both X and Y polarisation. The results as achieved on changing the dimensions tell us that the same design with some more optimal dimensions can also be used in multiple applications,. 
\bibliographystyle{IEEEtran}
\bibliography{example}

\end{document}